
\input phyzzx
%
%
\newcount\lemnumber   \lemnumber=0
\newcount\thnumber   \thnumber=0
\newcount\conumber   \conumber=0

\def\myeq{{\the\equanumber}}

\def\Lemma{\par\noindent\global\advance\lemnumber by 1
           {\bf Lemma\ (\chapterlabel\the\lemnumber)}}
\def\Corollary{\par\noindent\global\advance\conumber by 1
           {\bf Corollary\ (\chapterlabel\the\conumber)}}
\def\Theorem{\par\noindent\global\advance\thnumber by 1
           {\bf Theorem\ (\chapterlabel\the\thnumber)}}

%
%
\def\e{\adveq\eqno{\rm (\chapterlabel\the\equanumber)}}
\def\adveq{\global\advance\equanumber by 1}

\def\manyeq#1{\eqalign{#1}\e}

%
%
\font\tensl=cmsl10
\font\tenss=cmssq8 scaled\magstep1
\outer\def\quote{
   \begingroup\bigskip\vfill
   \def\endquote{\endgroup\eject}
    \def\par{\ifhmode\/\endgraf\fi}\obeylines
    \tenrm \let\tt=\twelvett
    \baselineskip=10pt \interlinepenalty=1000
    \leftskip=0pt plus 60pc minus \parindent \parfillskip=0pt
     \let\rm=\tenss \let\sl=\tensl \everypar{\sl}}
\def\from#1(#2){\smallskip\noindent\rm--- #1\unskip\enspace(#2)\bigskip}

\def\CALT{\address{Division of Physics, Mathematics
and Astronomy\break
Mail Code 452--48\break
California Institute of Technology\break
Pasadena, CA 91125}}

\def\r#1{$\lb \rm#1 \rb$}

%
%
\def\rarrow{\rightarrow}

\def\semidirect{\mathrel{\raise0.04cm\hbox{${\scriptscriptstyle |\!}$
\hskip-0.175cm}\times}}

\def\mod{\mathop{\rm mod}\nolimits}

\def\ref#1{$^{#1}$}

\def\half{{1\over2}}
\def\lb{\lbrack}
\def\rb{\rbrack}

\def\diam{{\hbox{\hskip-0.02in
\raise-0.126in\hbox{$\displaystyle\bigvee$}\hskip-0.241in
\raise0.099in\hbox{ $\displaystyle{\bigwedge}$}}}}

\Pubnum={}
\pubtype={Caltech preprint}
\date{October, 1994}
\titlepage
\title{Lattice Models and Generalized Rogers Ramanujan Identities}
\author{Doron Gepner\foot{On leave from: Department of Nuclear Physics,
Weizmann Institute of Science, Rehovot, Israel.}}
\CALT
\abstract
We revisit the solvable lattice models described by Andrews Baxter and
Forrester and their generalizations. The expressions for the local state
probabilities were shown to be related to characters of the minimal
models. We recompute these local state probabilities by a different method.
This yields generalized Rogers Ramanujan identities, some of which
recently conjectured by Kedem et al. Our method provides a proof for some
cases, as well as generating new such identities.
\endpage
\def\~{\sim}
 Several different systems in two dimensional physics appear to be
closely related. These include integrable $N=2$ systems, rational conformal
field theories (RCFT), solvable lattice models and integrable massive
field theories
(see, for example, ref. \REF\Found{D. Gepner, Caltech preprint, CALT-68-1825,
November 1992}\r\Found). There are other, rather
mysterious, examples of such connections. One is the appearance of the
characters of the fixed point conformal field theory in the expression for
the local state probabilities (LSP) on the lattice
\REF\Talk{D. Gepner, in ``Non perturbative methods in field theory'',
Irvine 1987, Conference proceedings, Nucl. Phys. A1 [PS] (1987) 105}
\REF\Date{E. Date, M. Jimbo, T. Miwa and M. Okado, Phys. Rev. B35 (1987) 2105}
\REF\LHP{E.  Date, M. Jimbo, A. Kuniba, T. Miwa and M. Okado, Nucl. Phys. B290
[FS20] (1987) 231}
\r{\Talk,\Date,\LHP}.
Presumably,
the explanation for this phenomenon lies in the context of the aforementioned
correspondence between lattice models and RCFT.
Another intriguing observation is the connection between generalized
Rogers--Ramanujan (GRR)
identities for the characters of RCFT, and the thermodynamic Bethe ansatz (TBA)
equations for the massive perturbation of the theory
\REF\RR{R. Kedem, T.R. Klassen, B.M. McCoy and E. Melzer, Phys. Lett. B304
(1993) 263; B.L. Feigin, T. Nakanishi and H. Ooguri, Int. J. Mod. Phys.
A7, Suppl. 1A (1992) 217; W. Nahm, A. Recknagel and M. Terhoeven, Mod. Phys.
Lett. A8 (1993) 1835; M. Tehoeven, Bonn preprint BONN-HE-92-36, hepth/9111120;
A. Kuniba, T. Nakanishi and J. Suzuki, Mod. Phys. Lett. A8 (1993) 1649}
\r\RR.

These observations remain largely unexplained. Our purpose here is to resolve
a little part of the puzzle. We do this by exploring yet another connection.
It is shown that the very same generalized Rogers--Ramanujan identities
naturally arise as an expression for the local state probabilities of the
corresponding lattice models. This closes a circle, as those LSP are
given in terms of the RCFT characters, which, in turn, can be expressed as
GRR. This provides a systematic   derivation for these identities, along with
a proof of their validity.

For the sake of concreteness the following picture is conjectured. There is
a one to one correspondence between RCFT, solvable lattice models, GRR, and
TBA. Our purpose here is to explore the connection between GRR and lattice
models. In this context we conjecture that every solvable lattice model leads
through the LSP to a GRR which is the character of the fixed point RCFT.
Indeed this is how Rogers--Ramanujan identities first arose in physics, in
the expressions for the LSP of the hard hexagon model \REF\Baxter{R.J. Baxter,
Exactly solved models in statistical mechanics, Academic Press, London, 1982}
\r\Baxter.
Specifically, we explore here the coset models $SU(2)_{k-m}\times SU(2)_m/
SU(2)_k$, which include the unitary minimal models, and
leave further exploration of the connection to future work.

Consider the Andrews--Baxter--Forrester model (ABF) \REF\ABF{G.E. Andrews,
R.J. Baxter and P.J. Forrester, J. Stat. Phys. 35 (1984) 193}
\r\ABF.
This model is described by a square lattice on which the state variables $l$
take the values $l=1,2,\ldots,k+1$ where $k$ is some integer which labels the
model. If $l_1$ and $l_2$ are two state variables sitting on the same bond
they must obey the admissibility condition $|l_1-l_2|=1$. This model is in
correspondence with the RCFT $SU(2)_k$ WZW model. The Boltzmann weights for
a face are given in ref. \r\ABF\ and depend on the temperature--like
parameter $p$ and the spectral parameter $u$. For $p=0$ the model is critical.
At the limit $p=0$, $u\rarrow i\infty$, the Boltzmann weights coincide
with the braiding matrices of $SU(2)_k$ \r\Found.

The calculation of the LSP proceeds by the corner transfer matrices method
invented by Baxter \r\Baxter. We shall concentrate here on Regime III of the
model where the critical point is described by the unitary minimal models
\REF\Huse{D.A. Huse, Phys. Rev. D30 (1984) 3908}
\r\Huse. The case $k=2$ is the Ising model. The $k$'th model corresponds to the
$k$'th minimal model, i.e., the coset, $SU(2)_{k-1}\times SU(2)_1/SU(2)_k$.
In regime III the following expression was found for the LSP \r\ABF.
The ground states are labeled by a pair of states, $a$ and $b$, such that
$|a-b|=1$ and the pattern of the ground state is a chess board:
$$l_{i,j}=\cases{a &$i+j= s \mod 2$,\cr b& $i+j=(1-s)\mod2$,\cr}\e$$
where $s=0$ or $1$. We can assume that $b=a+1$. The LSP is the probability
of finding the state $c$ at the $l_{0,0}$ site, for the $(a,b)$ ground state,
$$P(c|a,b)=\langle \delta(c,l_{0,0}) \rangle.\e$$
The corner transfer matrix calculation
reduces the problem to a one dimensional configuration sum \r\ABF,
$$P(c|a,b)={\chi_c^{k-1}(q,x) q^{-\nu}\phi(c|a,b) \over \chi_a^k(q,x)
\chi_1^1(q,x)},\e$$
where $\nu$ is some power given later, eq. (44).
$\chi_n^k(q,z)$ is the character of the $a$ field of the model $SU(2)_k$,
given by,
$$\chi_a^k(q,z)={\Theta_{a+1,k+2}(q,z)-\Theta_{-a-1,k+2}(q,z)\over
\Theta_{1,2}(q,z)-\Theta_{-1,2}(q,z) },\e$$
and
$$\Theta_{j,m}(q,z)=\sum_{\gamma\in Z+{j\over 2m}}
q^{m\gamma^2}z^{m\gamma},\e$$
and $q$ and $x$ are connected to the temperature--like parameter of the
lattice model, $p$,
$$p=e^{-\epsilon/(k+2)},\qquad x=e^{-4\pi^2/\epsilon},\qquad q=x^2.\e$$
The function $\phi(c|a,b)$ is the one dimensional configuration sum,
$$\phi(c|a,b)=\sum_{l_i} q^{\sum_{j=1}^n j|l_{j+2}-l_j|/4 },\e$$
where the first sum is over all admissible sequences $l_1\~l_2\~\ldots\~
l_{j+2}$,
such that $l_{n+1}=a$ and $l_{n+2}=b$, $l_1=c$, and the limit $n\rarrow\infty$
is taken. The configuration sum was computed in ref. \r\ABF\ and is
expressible by characters of the minimal models,
$$\phi(c|a,b)=q^\nu \chi_{a,c}(q),\e$$
where $\chi_{a,c}(q)$ is the character of the $(a,c)$ representation in the
$k$th minimal model,
$$\chi_{a,c}(q)={q^{-\Delta_{a,c}}\over \prod_{j=1}^\infty (1-q^j) }
\sum_{m\in Z} q^{\Delta_{a+2(k+1)m,c}}-q^{\Delta_{a+2(k+1)m,-c}},\e$$
and
$$\Delta_{a,c}={[a(k+2)-c(k+1)]^2-1\over4(k+1)(k+2)}.\e$$

Our purpose is to present an alternative calculation of the configuration
sum. This will provide us with the other side of the GRR identities.
For this purpose we define the following truncation of the configuration
sum,
$$G_r(l_r,l_{r+1})=\sum_{l_{r+2},l_{r+3},\ldots} q^{\sum_{j=l}^\infty
j|l_{j+2}-l_j|/4 },\e$$
where the sum is over all admissible sequences $l_r\~l_{r+1}\~l_{r+2}\~\ldots$,
and in such a way that the $(a,b)$ ground state is assumed, as before.
Evidently,
$$G_0(c-1,c)=G_0(c+1,c)=\phi(c|a,b).\e$$
The functions $G_r(t,y)$ obey the recursion relation,
$$G_r(t,y)=\sum_{z\atop z\~y} q^{|z-t|/4} G_{r+1}(y,z),\e$$
which is obtained by eliminating $l_r$. $G_r(t,y)$ also obeys the boundary
condition,
$$G_r(t,y)=\cases{1+O(q^r) & x=a {\rm and} y=b\cr O(q^r) &{\rm otherwise},}\e$$
for large $r$. The recursion relation eq. (13) along with the large $r$ limit,
eq. (14), uniquely defines $G_r$ and enables its calculation.

Next define the moments of $G_r$ by
$$G_r(t,y)=\sum_{n=0}^\infty q^{nr/2} a_n(t,y).\e$$
The moments $a_n(t,y)$ obey the recursion relation,
$$a_n(t,y)=\sum_{z\~y} q^{n/2-|z-t|/4} a_{n-\half|z-t|}(y,z),\e$$
which follows from eqs. (13,15), along with the initial value,
$$a_0(t,y)=\cases{1 & t=a {\rm and } y=b \cr
                  1& t=b {\rm and} y=a \cr
                  0& {\rm otherwise}.\cr}\e$$
Again, from the recursion relation, eq. (16) one may compute, in
principle at least, the moments. Our purpose is to solve these
recursion relations.

We may cast the recursions relation eq. (16) in a form that will be
more convenient, and also exhibits that it is indeed a recursion,
$$a_n(c-1,c)={q^{(n-1)/2} a_{n-1}(c,c+1)+q^{n-\half} a_{n-1}(c-1,c-2)\over
1-q^n},$$
$$a_n(c,c-1)={q^{n-\half} a_{n-1}(c,c+1)+q^{(n-1)/2}a_{n-1}(c-1,c-2)\over
1-q^n},\e$$
where $c=2,3,\ldots,k+1$ and we define $a_n(t,y)=0$ when $t$ or $y$ are out
of the range $1\leq t,y\leq k+1$.

For $k=2$, the Ising case, the recursion relations eq. (18) are solved
immediately, as they involve only one term. We easily find, for example,
for the $(a,b)=(1,2)$ phase,
$$a_n(2,1)={q^{\half n^2}\over (1-q)(1-q^2)\ldots (1-q^n)},\e$$
for even $n$, and $a_n(2,1)=0$ for odd $n$. Thus, we prove the identity,
$$\chi_{1,1}(q)=\sum_{n=0\atop n=0\mod2}^\infty {q^{\half n^2}\over
(1-q)(1-q^2)\ldots (1-q^n)},\e$$
which appears in the list of Slater
\REF\Slater{L.J. Slater, Proc. London Math. Soc. 54 (1953) 147}
\r\Slater\ of Rogers Ramanujan identities.

Let us introduce the notation
$$(q)_n=\prod_{j=1}^n (1-q^j),\e$$
and  the $q$--binomial coefficients,
\def\bk#1#2{{\left[{#1\atop#2}\right]}}
$$\bk m n={(m)_q\over (n)_q (m-n)_q},\e$$
if $m\geq n\geq0$, $m$ and $n$ integers, and $\bk m n=0$ otherwise.
The $q$--binomial coefficients, are polynomials in $q$, also called Gaussian
polynomials. This follows from the fundamental recurrences they obey
\REF\And{G.E. Andrews, The theory of partitions, (Addison--Wesley, London,
1976}\r\And,
$$\bk n m-\bk {n-1} m=q^{n-m}\bk {n-1} {m-1},\e$$
$$\bk n m-\bk {n-1} {m-1}=q^m \bk {n-1} m.\e$$

We found that the solutions to the recurrences $a_n(x,y)$ are closely related
to generalized Rogers--Ramanujan identities recently conjectured in ref.
\REF\Kedem{R. Kedem, T.R. Klassen, B.M. McCoy and E. Melzer, Phys. Lett. B307
(1993) 68}
\r\Kedem. To describe these, following \r\Kedem,
introduce the sum,
$$S_p(A,Q,u)=\sum_{m\in (2Z_{\geq0})^p+Q} q^{{1\over4} m C_n m-\half A m}
{1\over (q)_{m_1}}\prod_{s=2}^p \bk {\half(mI_n+u)_s} {m_s},\e$$
where $A,u\in Z^p$ are vectors, $Am=\sum_{s=1}^p A_s m_s$, $Q\in(Z_2)^p$.
$C_n$ is the cartan matrix of $A_n$ and $I_n=2-C_n$ is the incidence
matrix: $(I_n)_{ab}=\delta_{a,b+1}+\delta_{a,b-1}$.

We claim that $a_n(x,y)$ is related to the above sums.
Fix $p=k-1$. Denote by $M_{r,s}$ the sum
$$M_{r,s}=S_p(Q_{r,s},e_{p+2-s},e_r+e_{p+2-s}),\e$$
where
$$Q_{r,s}=(s-1)(e_1+e_2+\ldots+e_p)+(e_{r-1}+e_{r-3}+\ldots)+
(e_{n+3-s}+e_{n+5-s}+\ldots),\e$$
where $e_s$ is a unit vector, $(e_s)_x=\delta_{sx}$ and set $e_s=0$
for $s\notin \{1,2,\ldots,p\}$.
Then for the $(a,a+1)$ phase, $a_n(x,y)$ is given by the $m_1=n$
term in the sum for
$$a_n(c+1,c)=M_{a,c}\big|_{m_1=n},\e$$
$$a_n(c,c+1)=M_{p+2-a,p+2-c}\big|_{m_1=n},\e$$
$$a_n(p+1,p+2)=M_{a,p+2}\big|_{m_1=n}.\e$$

The proof proceeds by inserting the expressions for $a_n$, eqs. (28-30), into
the recursion relations, eqs. (18), and showing that they hold. The proof
is incomplete at the present as we can only show this for
$k=2,3$, and some of the recursion relations for higher $k$, but not all.

The $k=2$ case is the Ising model already mentioned. The recursion relations,
eq. (18) are solved immediately, as already discussed, and we find
expressions for
GRR which are already known and proved. So we proceed to the first non--trivial
case which is $k=3$. The proof of the recursion relations, eq. (18) is
a straight forwards application of the recurrences, eqs. (23,24).
We describe as a sample identity, the cases of $c=3$ in the second eq. (18),
for the $(2,1)$ phase,
$$a_n(3,2)(1-q^n)-a_{n-1}(3,4)q^{n-\half}=a_{n-1}(2,1)q^{(n-1)/2}.\e$$
We compute the l.h.s., substituting the expressions for $a_n$, eqs. (28-30),
$$\manyeq{
a_n(3,2)(1-q^n)-a_{n-1}(3,4)q^{n-\half}=\cr\sum_{{\rm odd} m}
{q^{\half(n^2+m^2-nm-m)}\over (q)_{n-1}}\left\{\bk{\half n+\half}{m}-q^m \bk
{\half n-\half}{m}\right\}=\cr{q^{\half(n^2+m^2-nm-m)}\over (q)_{n-1}}
\bk{\half n-\half}{m-1},}$$
where we used the recurrence, eq. (24). Computing the r.h.s of eqs. (31),
it immediately follows that it is the same as the r.h.s of eq. (32), with
the substitution of $m\rarrow m-1$, thus proving eq. (31).
The other recurrence relations are similarly easy to prove. It is also
evident that the $a_n$'s so defined obey the initial value,
eq. (17). This completes the proof that the $a_n$'s defined by eqs. (28-30)
indeed are the moments of $G_r$, eq. (15). From eqs. (8,15) it follows that
$$\sum_{n=0}^\infty a_n=\chi_{a,c},\e$$
where $\chi_{a,c}$ is the character of the minimal model, and we have proved
the GRR,
$$\chi_{a,c}=M_{a,c},\e$$

For $k>3$ we have only been able to prove some of the recursion relations.
Take, for example, the recurrences
$$a_n(2,1)=a_n(1,2)q^{n/2},\e$$
$$a_n(k,k+1)=a_n(k+1,k) q^{n/2}.\e$$
These follow as a straightforward application of the definition eqs. (28-30).
Similarly, the recurrences
$$a_n(1,2)=a_n(2,1)q^{n/2}+a_{n-1}(2,3)q^{(n-1)/2},\e$$
$$a_n(k+1,k)=a_n(k,k+1)q^{n/2}+a_{n-1}(k,k-1)q^{(n-1)/2},\e$$
follow immediately from the definition eqs. (28-30). This provides expression
for some of the characters if we assume that $a_n(2,1)$ and $a_n(k,k+1)$ are
indeed of the form eqs. (28-30). One can
also check that the initial value eq. (17)
holds. We have verified in many cases, by computer to high order, that indeed
$a_n$ is given by the expression eqs. (28-30). We believe
that upon further effort
the proof for $k>3$ can be completed along the lines described above.
\par
The results described here can be generalized to other models. Consider
the lattice models which correspond to IRF$(SU(2)_k,[m],[m])$,
where $[m]$ stands for the representation with highest weight $(m-1)\lambda$
where
$\lambda$ is the fundamental weight. The states of the lattice model are,
again, primary fields of SU$(2)_k$. The admissibility condition is given by
the fusion rule with respect to $[m]$, $a\~b$ if and only if
$$a=m+b\mod 2 {\ \rm and\ }|a-b|+1\leq m\leq \min(a+b-1,
2k-a-b+3).\e$$
The Boltzmann weights of the models can be obtained by the fusion procedure
and are described in ref. \REF\Datetwo{E. Date, M. Jimbo, T. Miwa and M. Okado,
Lett. Math. Phys. 12 (1986) 209}\r\Datetwo.

The local state probability of this model was computed in ref.
\r\LHP, and is given in terms of the one dimensional configuration
sum,
$$\phi(c|a,b)=\sum_{l_i} q^{\sum_{j=1}^n j|l_{j+2}-l_j|/4},\e$$
where $n\rarrow\infty$, $l_1=c$, $l_{n+1}=a$, $l_{n+2}=b$, and the sum is
over admissible sequences, $l_1\~l_2\~l_3\~\ldots\~l_{n+2}$.
Notice that this configuration sum is identical to what was found before for
$m=2$, eq. (7), except for the admissibility condition which changes.
The critical theory of this lattice model is the coset RCFT
$O=$ $SU(2)_{k-m+1}\times SU(2)_{m-1}\over SU(2)_k$.
It is found in ref. \r\LHP\ that the configuration sum $\phi(c|a,b)$ is
given by the characters of this RCFT which are the branching functions
associated to this coset.
Define,
$$r=\half(a+b-m+1),\qquad s=\half(a-b+m+1),\e$$
the configuration sum is given by,
$$\phi(c|a,b)=q^\nu c_{rsc}(q),\e$$
where $c_{rsc}(q)$ is the character of the RCFT $O$ defined by,
$$\chi_r^{k-m+1}(q,z) \chi_s^{m-1}(q,z)=\sum_a c_{rsa}(q)\chi_a^k(q,z),\e$$
where $\chi_r^m(q,z)$ is the character of SU$(2)$ at level $k$ and isospin
$r$, defined by eq. (4). The power $\nu$ is given by
$$\nu=\half(b-c)+ \gamma(r,s,c),\e$$
where
$$\gamma(r,s,c)={r^2\over 4(k-m+1)}+{s^2\over 4(m-1)}-{c^2\over 4k}-{1\over8}
.\e$$
The local state probability in regime III
is given by
$$P(a|b,c)={\chi_a^k(x^2,x) c_{rsa}(q)\over \chi_r^{k-m+1}(x^2,x)\chi_s^
{m-1}(x^2,x)},\e$$
where $q$ and $x$ are defined in eq. (6).

We can again try to calculate the configuration sum, $\phi(a|b,c)$ in another
way by defining $G_r(l_r,l_{r+1})$ as in eq. (11),
$$G_r(l_r,l_{r+1})=\sum_{l_{r+2},l_{r+3},\ldots} q^{\sum_{j=l}^\infty j
|l_{j+2}-l_j|/4},\e$$
where the sum is over all admissible sequences $l_r\~l_{r+1}\~l_{r+2}\~\ldots$,
and is taken in the $(a,b)$ phase. As before,
$$G_0(d,c)=\phi(c|a,b),\e$$
for all $d$ such that $d\~c$. We have the same recursion relation for
$G_r(t,y)$
as before, eq. (13). We define the moments of $G_r(t,y)$, $a_n(t,y)$, as
before, eq. (15), and they obey a similar recursion relation,
$$a_n(t,y)=\sum_{z\~y} q^{n/2-|z-t|/4} a_{n-\half|z-t|}(y,z),\e$$
along with the initial value, eq. (17).

The complete solution for the moments $a_n(t,y)$ is not known. However,
we can obtain parts of the solution, and through that new GRR identities.
Our starting point is the conjecture proposed in ref. \r\Kedem\ for the
character of
the identity in the RCFT $O$. This is given by the expression,
$$M=\sum_{l_1,l_2,\ldots,l_n \in (2Z_{\geq0})^n} {q^{{1\over4}lC_nl}\over
(q)_{l_{m-1}}}\prod_{a=1\atop a\neq m-1}^n \bk{\half (l_{a-1}+l_{a+1})}
{l_a},\e$$
where $C_n$ is the Cartan matrix of $A_n$.
We find that
$a_n(m,1)$ in the $(1,m)$ phase is given by the $l_{m-1}=n$'th term of eq.
(50),
$$a_n(m,1)=M\bigg|_{l_{m-1}=n},\e$$
for even $n$, and is zero for odd $n$.
This is consistent with the fact that the identity character is given by
$$c_{1,1,1}(q)=\sum_{n=0}^\infty a_n=M.\e$$
We checked this in various examples by computer to high order. Accepting this
conjecture, we can immediately find new GRR by utilizing the recursion
relations eq. (49). From the recursion relation,
$$a_n(m,1)=a_n(1,m) q^{n/2},\e$$
we get an expression for $a_n(1,m)$,
$$a_n(1,m)=q^{-n/2} M\big|_{l_{m-1}=n},\e$$
which leads to the character identity,
$$c_{1,1,m}=\sum_{l_1,l_2,\ldots,l_n\in (2Z_{\geq0})^n} {q^{{1\over4}lC_nl-
\half l_{m-1}} \over (q)_{l_{m-1}}} \prod_{a=1\atop a\neq m-1}^n
\bk {\half(l_{a-1}+l_{a+1})}{l_a}.\e$$

As an example, we have computed characters for $m=3$ and
$k=4$. This is the second minimal model for the super-Virasoro algebra.
The states are labeled by $a\in\{1,2,\ldots,5\}$. We find for the
moments $a_n(x,y)$ in the $(3,1)+(1,3)$ phase,
$$a_n(3,1)=\sum_{m_3 {\rm\ even},m_1} {q^{\half(m_1^2+n^2+m_3^2-nm_1-nm_3)}
\over (q)_n} \bk{\half n}{m_1}\bk{\half n}{m_3},\e$$
for even $n$; $a_n(3,1)=0$ for odd $n$.
$$a_n(1,3)=\sum_{m_3 {\rm\ even},m_1} {q^{\half(m_1^2+n^2+m_3^2-nm_1-nm_3-n)}
\over (q)_n} \bk{\half n}{m_1}\bk{\half n}{m_3},\e$$
for even $n$; $a_n(1,3)=0$ for odd $n$.
$$a_n(3,5)=\sum_{m_3 {\rm \ odd},m_1}{q^{\half(m_1^2+n^2+m_3^2-nm_1-nm_3)}
\over (q)_n} \bk{\half n}{m_1}\bk{\half n}{m_3},\e$$
for even $n$. $a_n(3,5)=0$ for odd $n$.
$$a_n(5,3)=\sum_{m_3 {\rm\ odd},m_1} {q^{\half(m_1^2+n^2+m_3^2-nm_1-nm_2-n)}
\over (q)_n}\bk{\half n}{m_1}\bk{\half n}{m_3},\e$$
for even $n$.
$$a_n(3,3)=\sum_{m_1,m_3} {q^{\half(m_1^2+n^2+m_3^2-nm_1-nm_3-n+m_1+m_3)}\over
(1-q^{n/2}) (q)_{n-1}}\bk{\half(n-1)}{m_1}\bk{\half(n-1)}{m_3},\e$$
for odd $n$.
In the $(3,3)$ phase we find,
$$a_n(3,1)=\sum_{m_1,m_3 {\rm\ odd}} {q^{\half(m_1^2+n^2+m_3^2-nm_1-nm_3)}
\over (q)_n} \bk{\half (n+1)}{m_1}\bk{\half (n+1)}{m_3},\e$$
for odd $n$; $a_n(3,1)=0$ for even $n$.
$$a_n(1,3)=\sum_{m_1,m_3 {\rm\ odd}} {q^{\half(m_1^2+n^2+m_3^2-nm_1-nm_3-n)}
\over (q)_n} \bk{\half (n+1)}{m_1}\bk{\half (n+1)}{m_3},\e$$
for odd $n$; $a_n(1,3)=0$ for even $n$.
$$a_n(3,3)=\sum_{m_1,m_3 {\rm\ odd}} {2q^{\half(m_1^2+n^2+m_3^2-nm_1-nm_3
-n+m_1+m_3)}
\over (1-q^{n/2})(q)_{n-1}} \bk{\half n}{m_1}\bk{\half n}{m_3},\e$$
for even $n$; $a_0(3,3)=1$; $a_n(3,3)=0$ for odd $n$.
Using the symmetry property $a_n(x,y)$ in the $(a,b)$ phase is the same
as $a_n(k+2-x,k+2-y)$ in the $(k+2-a,k+2-b)$ phase, we get all the
moments of the configuration sum for $x,y,a,b$ even (the even sector).

We find nice expressions for most of the sums in the odd sector, as well.
In the $(2,2)$ ground state we have,
$$a_n(2,4)=\sum_{m_1{\rm\ even},m_3{\rm\ odd}} {q^{\half(m_1^2+n^2+m_3^2-nm_1
-nm_3-m_1)}\over (q)_n}\bk{\half(n+1)}{m_1}\bk{\half(n+1)}{m_3},\e$$
for even $n$.
$$a_n(2,2)=\sum_{m_3{\rm\ odd},m_1{\rm\ even}} {q^{\half(m_1^2+n^2+m_3^2-
nm_1-nm_3+m_3-n)}\over (1-q^{n/2})(q)_{n-1}}\bk{\half n}{m_1}\bk{\half n}
{m_3},\e$$
for even $n$.
In the $(2,4)+(4,2)$ ground state we find,
$$a_n(2,2)=\sum_{m_2 {\rm\ odd},m_1} {q^{\half(m_1^2+n^2+m_3^2-nm_1-nm_3-m_1)}
\over (q)_n} \bk{\half(n+1)}{m_1}\bk{\half(n+1)}{m_3},\e$$
for odd $n$.
$$a_n(2,4)=\sum_{m_3{\rm\ odd},m_1} {q^{\half(m_1^2+n^2+m_3^2-nm_1-nm_3+
n-m_1-m_3)}\over
(1+q^{(n+1)/2})(q)_n}\bk{\half n+1}{m_1}\bk{\half n+1}{m_3},\e$$
for even $n$.

By summing over $a_n(t,y)$ we find new GRR character identities.
Since $G_0(t,y)=\sum_{n=0}^\infty a_n(t,y)=c_{rsa}(q)$, we get the identities,
$$c_{111}+c_{131}=G_0(3,1),\e$$
$$c_{113}+c_{133}=G_0(1,3)=G_0(3,3)=G_0(5,3),\e$$
$$c_{115}+c_{135}=G_0(3,5),\e$$
in the $(1,3)$ phase. In the $(3,3)$ phase we find,
$$c_{221}=G_0(3,1),\e$$
$$c_{223}=G_0(1,3)=G_0(3,3).\e$$
In the $(2,2)$ phase we find,
$$c_{124}=G_0(2,4),\e$$
$$c_{122}=G_0(2,2).\e$$
In the $(2,4)$ phase we find
$$c_{212}+c_{232}=G_0(2,2)=G_0(2,4).\e$$
This completes the description of new GRR associated with
this theory. We expect that this can be generalized to all $m$ and all
$k$.

We described here the appearance of generalized Rogers--Ramanujan identities
in the expressions for the local state probability in solvable lattice models.
We believe that this correspondence is quite general and could be extended to
many other such models. It is hoped that this work will be of help in
understanding solvable lattice models, their relationship to RCFT, and
generalized Rogers--Ramanujan identities.
\ACK
The author gratefully acknowledges the hospitality of the theory division of
CERN where part of this work was done, and thanks J. Schwarz for a critical
reading of the manuscript.
\refout
\bye